\documentclass{proceedingsM}
\usepackage{amsmath,amssymb}
\usepackage[numbers,sort&compress]{natbib}
\usepackage{color}

\title{On concepts of mathematical physics for modelling signals in axons}

\author{Tanel Peets$^1$, Kert Tamm$^1$, J\"uri Engelbrecht$^{1,2}$}
\address{1. Department of Cybernetics, School of Science, Tallinn University of Technology, Ehitajate 5, Tallinn, 19086, Estonia}
\address{2. Estonian Academy of Sciences, Kohtu 6, 10130 Tallinn, Estonia} 
\email{tanelp@ioc.ee, kert@ioc.ee, je@ioc.ee}

\abstract{In this short paper, the results of the paper by Drab et al. (Eur. Phys. J. E (2022) 45:79) are described in the framework of wave mechanics and mathematical physics based on common understandings.  The attention is focused on properties of Boussinesq-type equations, solitons, and peakons. These concepts  are supported by several experimental observations.}

\begin{document}
\maketitle

\section{Introduction}

Modelling of signal 
propagation in nerve fibres is of the upmost importance. It is related not only to understanding the processes in single nerves, but is related to cognitive processes and neural information. There are many objective measurements of signal starting from the action potential (AP) up to other effects like mechanical deformation of biomembrane and temperature changes. The modelling of such a complex process needs interdisciplinary studies and the notions from different fields must be properly used.  The present short note is based on  the recent paper by Drab et al. \cite{Drab2022} that needs  a proper framework for being comparable to other studies. A shorter version of the ideas presented here is published in Eur. Phys. J. E \cite{Peets2023a}.

\section{Boussinesq paradigm and solitary waves}

The Boussinesq-type equation is an important extension of the classical wave equation and is characterised by the following properties \cite{Christov2007}:
\begin{itemize}
	\item[(i)] bi-directionality of waves (due to the second order wave operator);
	\item[(ii)] nonlinearity of any order;
	\item[(iii)] dispersion modelled by fourth or higher order spatial and/or time derivatives. 
\end{itemize}
  The general form of such an equation is
\begin{equation}
\label{BOEq}
	u_{tt}-c_0^2u_{xx}-\left(\frac{\mathrm{d} F(u)}{\mathrm{d} u}\right)_{xx}
	=(\beta_1u_{tt}-\beta_2u_{xx})_{xx},
\end{equation}
where $F(u)$ is a polynomial of second (or higher) degree, $c_0$ is velocity and $\beta_i$ are coefficients characterising dispersion. This equation models the so-called Boussinesq paradigm \cite{Christov2007}.  The nonlinearities in Eq.~\eqref{BOEq} can 	 be modelled differently. For example, Heimburg and Jackson \cite{Heimburg2005} have modelled nonlinearities in biomembrane as
\begin{equation}
	\label{HJAssump}
	c^2=c_0^2+pu+qu^2,
\end{equation}
where $c_0$ is the velocity of sound in unperturbed state, $u$       is the change in density, and coefficients $p$, $q$ are determined from the experiments.

Equation~\eqref{BOEq} is the form that was initially derived by Boussinesq, but in an attempt to obtain analytical solutions, a fourth order mixed partial derivative term was dropped \cite{Christov2007} and to this date it is common that Eq.~\eqref{BOEq} with $\beta_1=0$ is usually presented in literature and is sometimes called a `good' Boussinesq equation. A `bad' Booussinesq equation has a positive sign in front of $\beta_2$ and is mathematically improper as short wavelengths lead to imaginary frequencies \cite{Christov2007}.

\section{Solitary wave solutions}

The coexistence of nonlinearity and dispersion in Eq.~\eqref{BOEq} can give rise to solitary wave solutions for suitable combination of parameters. Simple explanation for their existence is the balance of dispersion and nonlinearity -- dispersion leads to oscillatory behaviour of an initial excitation and nonlinearity leads to the formation of discontinuities. 

A standard technique for finding steady state solutions for Eq.~\eqref{BOEq} is to seek travelling wave solutions in the form
\begin{equation}
	V=V(\xi), \quad \xi=X-cT,
\end{equation}
where $V$ is some function and $c$ is the velocity of a moving frame. By this transformation a partial differential equation (PDE) is transformed into an ordinary differential equation (ODE) and the existence of solutions can be analysed in a phase space \cite{Ablowitz2011,Engelbrecht2017}. Due to the existence of nonlinearities in Eq.~\eqref{BOEq}, the fixed points on a phase space can be stable or unstable depending on the choice of parameters. For the existence of a solitary wave there has to be a saddle point and a homoclinic orbit around a centre \cite{Ablowitz2011,Engelbrecht2017}. The homoclinic orbit is a trajectory in phase space that connects a saddle point to itself.

Note that instead of a term `soliton', a term `solitary wave' is used in physics if  all conditions for a `soliton' are not satisfied. The term `soliton' was originally introduced  by Zabusky and Kruskal in their 1965 seminal paper \cite{Zabusky1965} and it involves more than just being localised in space, like, for example, the recurrence effect. 

In general, there are two classical definitions for a soliton: (a) a soliton can be described as a stable particle-like state in a nonlinear system \cite{SaNWE1982}, (b) another way is defining a soliton through its properties as a wave in the nonlinear environment that has (i) a stable form, (ii) is localised in space, (iii) restores its speed and structure after interaction with another soliton \cite{Ablowitz2011}. In this context only fulfilling the first two conditions is not sufficient to declare a localised pulse as a soliton.

 Also, it is important to check whether the interactions of solitary waves are elastic or not. If the interactions are not elastic, then the term `solitary wave' should be used instead of `soliton'. Unfortunately, in many studies  the conditions for existing solitons are not checked.  More on properties of solitons can be found in \cite{Ablowitz2011}. 



\section{Solitary waves in biomembranes}

It has been known for many decades  that the propagation of an AP is accompanied by a mechanical wave. This has been experimentally measured many times since the 1940's (see the review by J\'erusalem et al.  \cite{Jerusalem2019a} and references therein). First serious attempt to model the mechanical effects in biomembrane was done by Heimburg and Jackson \cite{Heimburg2005} in 2005. 
Their starting point is a classical wave equation for a density wave in biomembrane with an assumption that the wave velocity is dependent on the longitudinal density change (Eq.~\eqref{HJAssump}) and an \emph{ad hoc} term in the form of a fourth-order derivative with respect to the space coordinate was added in order to account for experimentally measured anomalous dispersion, i.e., higher frequencies travel at greater velocities \cite{Heimburg2005}. Since solitary waves emerge due to the balance of dispersive and nonlinear effects, then the dispersive term can be used to control the width of a pulse. Later the Heimburg-Jackson (HJ) model was improved by Engelbrecht et al. \cite{Engelbrecht2015} with addition of a dispersive term $u_{xxtt}$. While the dispersive term $u_{xxxx}$ models the elastic effects of a biomembrane, the dispersive term $u_{xxtt}$ models the inertial effects of a biomembrane. Both dispersive terms control the width of the soliton \cite{Engelbrecht2017}. 

In \cite{Drab2022} Drab et al.  derive an equation similar to the HJ model.  Their starting point is a Lagrangian for a transverse displacement and in the first linear approximation the Lagrangian $L$ is taken as
\begin{equation}
	L=\frac{1}{2}(\mu u^2_t-k_au_x^2-k_b u^2_{xx}),
\end{equation}
where $\mu$ is linear density, $k_a$ and $k_b$ are compression and bending modulus respectively.  By making use of the standard techniques, the dispersive term $u_{xxxx}$ arises naturally and does not need to be introduced \emph{ad hoc.}  


Nonlinear model (Eq.~(7) in \cite{Drab2022}) is derived by the authors assuming that the compression modulus $k_a$ is a function of transverse displacement:
\begin{equation}
	k_a(u)=a(a-P)^2+Q,
\end{equation}
where $a$, $P$, $Q$ are expansion coefficients determined from the experiments. The derived equation is similar to the HJ equation including the nonlinear terms $uu_x^2$, $u^2_x$, $u^2u_{xx}$ and $uu_{xx}$ along with the dispersive term $u_{xxxx}$. 

The basic studies of wave mechanics describe several models for nonlinear longitudinal  and transverse waves \cite{Bland1988} and here the question on how coefficients $a$, $P$ and $Q$ are related to the lipid structure of the biomembrane is not answered by physical considerations. 

Another question is the dispersion of transverse waves. Heimburg  and Jackson \cite{Heimburg2005} have added and \emph{ad hoc} dispersive term for the model of longitudinal waves. Engelbrecht et al. \cite{Engelbrecht2017} have shown that two terms are needed for describing the dispersion in a lipid bi-layer caused by elasticity and inertia of the lipid structure. What is the motivation to use the similar dispersive term like for a model of longitudinal waves also for transverse waves -- this must be clearly explained. Note that Selim \cite{Selim2019} has derived a model of transverse waves for a carbon nanotube directly from the principle of nonlocal continuum mechanics. The derived linear equation of motion has two dispersive terms like the improved Heimburg-Jackson  equation \cite{Engelbrecht2015}.

\section{Unipolarity of solutions}

The HJ equation and Eq.~(7) in \cite{Drab2022} have solitary wave solutions that are unipolar and the authors conclude that their model describes the same effect as the HJ model. However, one needs to keep in mind the HJ model is derived for the longitudinal density change, Drab et al. \cite{Drab2022} derive a model for the transverse displacement. The difference is that the longitudinal density change in biomembrane can be modelled by a unipolar pulse, the measured transverse displacement is not unipolar \cite{Tasaki1988,Terakawa1985, Engelbrecht2021a}. Moreover, it is known from continuum mechanics that the longitudinal and transverse displacements are coupled. In case of theory of rods, the transverse displacement $w$ is coupled to the longitudinal displacement as $w\propto -u_x$. This means that in case of unipolar transverse displacement, a kink-type longitudinal wave, i.e. permanent density change, should be present. 

\section{On peakons}

In Appendix A, the authors attempt to find a solitary wave solution for the linear dispersive wave equation (Eq.~(2) in \cite{Drab2022}) and they call it a peakon. Peakons are an interesting case of solutions that were first recognised in case of water waves by Camassa and Holm \cite{Camassa1993}.
However, it is common knowledge that peakons and other solitary wave solutions exist only in case of nonlinear dispersive hyperbolic partial differential equations. In phase space, a solitary wave solutions exist if there is a homoclinic orbit connecting a saddle point around a centre. A peakon solution exists when heteroclinic connection in phase space is present \cite{Liu2002a}. In case of a linear wave equation only a saddle point exists in phase space and solitary wave solutions are certainly not possible. By the first glance Eq.~(25) in \cite{Drab2022} looks very similar to a peakon solution derived by Camassa and Holm \cite{Camassa1993}:
\begin{equation}
	u(x,t)=c\,e^{-\lvert x-ct \rvert},
\end{equation}
where $c$ is the wave speed and $u$ is the fluid velocity. Drab et al. \cite{Drab2022} derived a solution 
\begin{equation}
	u(x,t)=c_1\,e^{-\lvert x-vt\rvert\sqrt{(c^2-v^2)/h}},
\end{equation}
where $c_1$ is the initial condition and $v$ is the speed of a moving frame. This solution denotes a arbitrary wave (determined by initial condition $c_1$) in a moving frame. In case of $c=v$, the solution is simply equal to the initial condition.

\section*{Acknowledgments}

The authors are supported by the Estonian Research Council (PRG 1227).



\end{document}